\documentclass[11pt]{article}

\usepackage{graphicx}
\usepackage[latin1]{inputenc}
\usepackage{natbib}
\usepackage{color}
\usepackage{amsbsy}
\usepackage{bm}
\usepackage{titlesec}
\usepackage[english]{babel}
\usepackage[normalem]{ulem} 

\usepackage{amssymb,amsmath,amsthm,amscd,eucal,mathrsfs,dsfont,verbatim}

\newcommand{\Cov}{\textrm{Cov}}
\newcommand{\Var}{\textrm{Var}}

\renewcommand{\leq}{\leqslant}
\renewcommand{\geq}{\geqslant}

\newcommand{\sphere}{\mathbb S}

\newcommand{\real}{\mathds{R}}
\newcommand{\rn}{{{\mathds R}^n}}

\newcommand{\rd}{{{\mathds R}^d}}

\newtheorem{theorem}{Theorem}

\theoremstyle{definition}

\def\urltilda{\kern -.15em\lower .7ex\hbox{\~{}}\kern .04em}

\begin{document}
\bibliographystyle{jecol}

\baselineskip28pt
\phantom{MEE}

\vspace{3cm}
\begin{center}
{\huge \bf Validity of covariance models for the analysis of geographical variation}\\
\bigskip
\bigskip
{\Large Gilles Guillot\footnote{Applied Mathematics and Computer
    Science Department,
 Technical University of Denmark,
Richard Petersens Plads, Bygning 321, 2800 Lyngby,  Denmark. {\tt email: gigu@dtu.dk}.},
Ren\'e L.\ Schilling\footnote{Technische Universit\"at Dresden, Institut f\"ur Mathematische
  Stochastik, 01062 Dresden, Germany. {\tt email:
    rene.schilling@tu-dresden.de}.},
Emilio Porcu\footnote{Universidad
  Federico Santa Maria, Department of Mathematics, Valparaiso, Chile.
  {\tt email: emilio.porcu@uv.cl}.}
and Moreno
Bevilacqua\footnote{Universidad de Valparaiso, Department of
  Statistics, Valparaiso, Chile.  {\tt email: moreno.bevilacqua@uv.cl}.}
}

\end{center}

\vspace{3cm}

\newpage
\topmargin30pt
\centerline{\textbf{Summary}}
{
\begin{enumerate}
\item Due to the availability of large molecular data-sets, covariance
  models are increasingly
used to describe the   structure of genetic variation as an
alternative to more heavily parametrised biological models.
\item We focus here on a class of parametric covariance models that
  received sustained attention lately and show that the  conditions
  under which they are valid mathematical models have been overlooked
  so far.
\item We provide rigorous results for the construction of valid
  covariance models in this family.
\item We also outline how to construct alternative
  covariance models for the analysis of geographical variation that are
  both mathematically well behaved and easily implementable.
\end{enumerate}
}

\bigskip
\bigskip


\newpage

\topmargin-45pt

\section{Background}
The spatial auto-covariance  function quantifies the linear 
statistical dependence between observations of a variable 
measured repeatedly across space. It has long been considered
a useful tool in studies that involve spatially structured variables in ecology and evolution.
It is indeed used at an exploratory and descriptive stage to identify
characteristic scales  of variation of the data
\citep{Levin92,Jackson93,Perry02}, it plays a central role in methods
for spatial
prediction \citep{Robertson87,Liebhold93,Hay09} and it is also involved in
regression-type analyses where an explicit spatial model
is used as a way to avoid confounding effects due to spatial
auto-correlation \citep{DinizFilho03,Diggle07b,Rahbek07}.
In recent years, the advent of new genotyping techniques has triggered a flood of
population genetics data in ecology. These data-sets are large and of ever
increasing sizes, therefore they can not be handled with heavily
parametrised models.
This situation has rekindled interest in approaches based on the
covariance structure of data. 
Indeed, although of rather descriptive nature compared to biologically
explicit models, covariance-based approaches
can capture characteristic scales in a parcimonious way 
and offer computationally efficient ways to recover information about evolutionary
processes.

\begin{figure}[h]
\vspace{.5cm}
\begin{tabular}{cc}
\vspace{.001cm}\includegraphics[width=7.8cm]{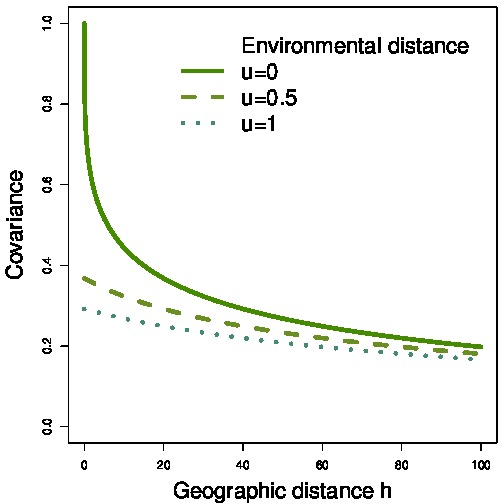} & \vspace{.001cm}\includegraphics[width=7.8cm]{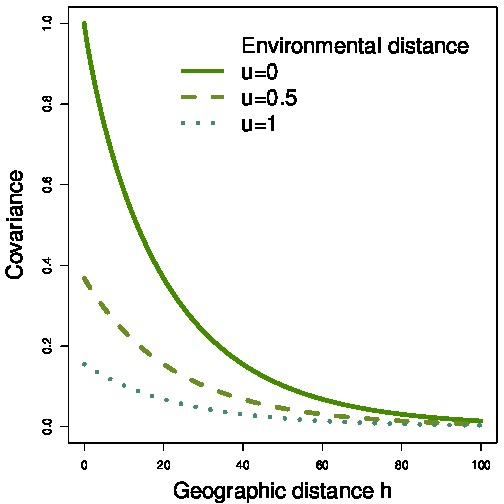}
\end{tabular}
\caption{Cross-sections of the BRC covariance function
$C(h,u) = 1/\alpha_0 \exp\left[ - \left( \alpha_G h +
    \alpha_E u \right)^{\alpha_2}\right]$ with $\alpha_0=1$,
  $\alpha_G=1/20$ and $\alpha_E= 2$.
Left panel $\alpha_2=0.3$, right panel: $\alpha_2=0.9$.}\label{fig:plotcov_BRC}
\end{figure}

In a recent paper, \citet{Bradburd13} introduced a method to
quantify the relative effects of geographic and ecological isolation on
genetic differentiation, making it possible to investigate the role of these
two factors on migration and gene flow.
In the model considered, a sample of individuals from a locality is indexed by its geographic coordinates $x$
and a  quantitative environmental
variable $e$. The frequency of an allele $f(x,e)$ is assumed to be
a suitable transform of a Gaussian random variable $y(x,e)$.
One of the key assumptions of the method is that the covariance structure of $y(x,e)$
is of the form:
\begin{equation}\label{eq:BRC}
\Cov\left[y(x,e),y(x',e') \right]
= C(h,u)
= \frac{1}{\alpha_0} \exp\left[ - \left( \alpha_G h +
    \alpha_E u \right)^{\alpha_2}\right]
\end{equation}
hereafter referred to as BRC model.
In the formula \eqref{eq:BRC}, $h$ and $u$ denote the geographic  and environmental distances between
samples indexed by  $(x,e)$ and $(x',e')$. The parameters $\alpha_0, \alpha_G, \alpha_E$ and $\alpha_2$ are
positive numbers which have to be inferred from the data.
The ratio $\alpha_E/\alpha_G$  can be interpreted as
the geographic distance equivalent to a unit environmental distance.
Plots of the spatial margins of this covariance function are shown in
Figure~\ref{fig:plotcov_BRC}.

This model is an extension of a simpler model which is known as the stable (or
powered exponential)
covariance \citep{Chiles99,Diggle07} and defined as
\begin{equation}\label{eq:stable}
    K(h) = \frac{1}{\alpha_0} \exp[-(\alpha_G h )^{\alpha_2}].
\end{equation}
The latter has been used by \citet{Wasser04,Wasser07} and
\citet{Rundel13} to perform spatial continuous assignment from genetic
data,  by \citet{Novembre08} to investigate the pattern in principal
components of geographically structured population genetics data and by \citet{Guillot09a} to
assess the effect of spatial sampling on the performances of
spatial clustering methods.

The use of spatial covariance functions has a long tradition in statistics and the model and method 
proposed by  \citet{Bradburd13}
can be advocated as well grounded alternative to the widely criticized partial Mantel test \citep{Guillot13}. 
The stable covariance and the BRC extension in particular 
can capture complex patterns of genetic
variation, yet they depend  on a small number of
parameters; as such, they are potentially
useful tools for modelling spatial variation in ecology and evolution.
Despite its apparent simplicity, this family of covariance functions
contains a subtle, but crucial, difficulty: not every function is a covariance function.\\
In this note, we first clarify what is involved in the
specification of a covariance model and show that some of the models
used earlier are not valid. Then, standing on a firm
mathematical footing, we provide results on the
range of validity of the models defined above and
outline  alternative way of constructing valid covariance models.
We conclude by discussing implications of our findings for earlier works.

\clearpage
\section{A covariance model must be a positive-definite function}

\subsection{Theoretical aspects}
Considering values $y(x_i,e_i)$ at $n$ locations  in the
geographical $\times$ environmental domain,
the variance of a weighted sum can be written
\begin{equation}\label{eq:var_cl}
    \Var \bigg[\sum_{i=1}^n \lambda_i y(x_i,e_i)\bigg]
    = \sum_{i=1}^n \sum_{j=1}^n
    \lambda_i \lambda_j \Cov[y(x_i,e_i),y(x_j,e_j)]
\end{equation}
and it is $\geq0$ 
for any combination of weights $\lambda_1,
\ldots, \lambda_n$.
Using a mathematical phrasing: the covariance function
$\Cov[y(x_i,e_i),y(x_j,e_j)]$ is a 
positive-definite function.
Consequently, if
one intends to use a certain covariance function considered suitable
(e.g. for modelling
or computational reasons), one has to make sure that it is
positive-definite, i.e. the expression in Equation~\eqref{eq:var_cl} has to be non-negative.


A scientist using a covariance model without this
property is likely to face negative variances
and undefined probability densities when embedding this covariance
function into a Gaussian model.
This would also thwart any simulation algorithm based on the Choleski
decomposition.
In other words, this model would make little sense.
It is therefore important to know whether the functions $C$ and $K$
defined by Equations~(\ref{eq:BRC}-\ref{eq:stable}) are
valid in this respect, or in mathematical parlance: When are $C$
and $K$ positive-definite functions?
This question has been overlooked so far and holds a number of subtleties, among others the fact that
(i) validity in a certain dimension does not imply validity in higher
dimensions, and importantly here,
(ii) the answer depends on the way
distances are measured (for example Euclidean in the plan vs.\ geodesic
distance on the earth's surface).

\subsection{A worked example: spatial prediction of tree abundance data with an invalid covariance model}
\begin{figure}[h]
\vspace{-.7cm}
\begin{tabular}{c}
\includegraphics[width=7.5cm]{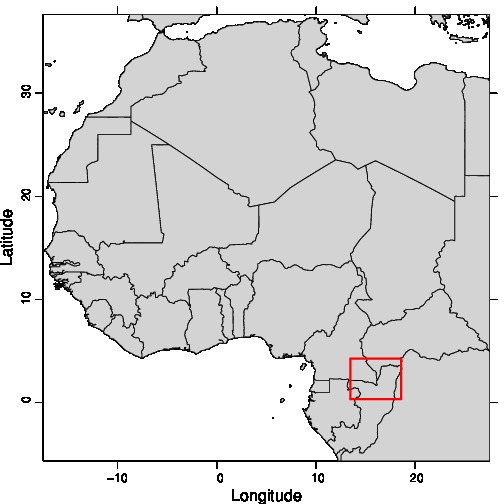} \includegraphics[width=8.5cm]{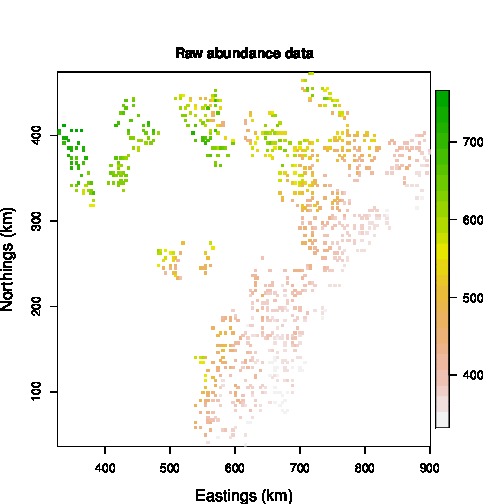} \\
\end{tabular}
\caption{Study area and tree abundance data in the tropical forest of the
  Congo-Basin in thousand 8km$\times$8km plots. }\label{fig:africa}
\end{figure}

\begin{figure}[h]
\centerline{\includegraphics[width=10cm]{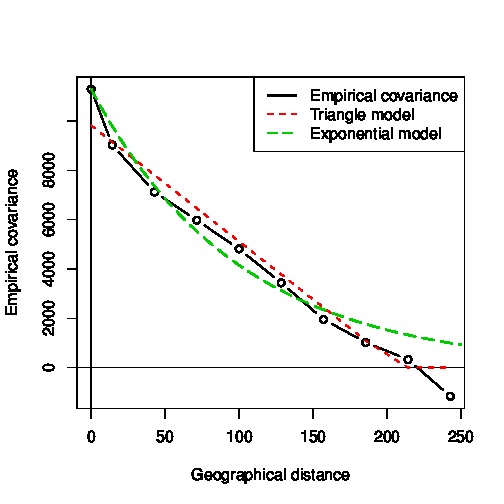}}
\caption{Empirical and theoretical covariances for the tree abundance data.
Distances are in kilometers.}\label{fig:cov}
\end{figure}

\begin{figure}[h]
\hspace{-.7cm}\includegraphics[width=17cm]{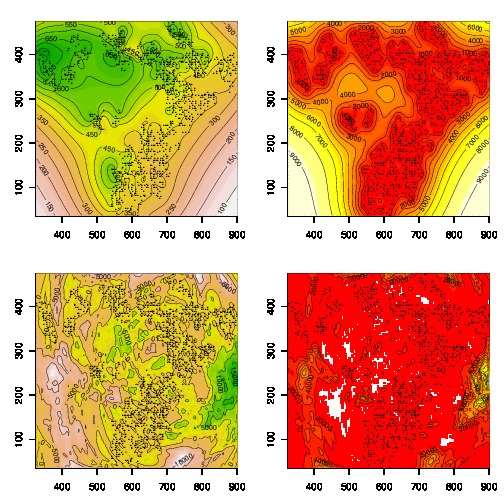}
\caption{Spatial prediction of tree abundance data in the tropical forest of the
  Congo-Basin.
Top: computations with an exponential covariance function.
Bottom: computations with a triangular function.
Left: abundance map obtained by simple kriging.
Right: kriging variance (white areas in bottom right panel correspond
to negative kriging variances).
Eastings and Northings in kilometers.}\label{fig:kriging}
\end{figure}

We illustrate some of the  consequences of using an invalid covariance model on  abundance data
for a tree genus  in the  moist forest of the Congo basin.
These data have been published by \citet{Mortier14}  and made publicly
available via the R package SCGLR.
The variable considered here  consists of abundance in thousand 8km by 8 km plots.
The location of sampling sites and abundance data are shown in Figure \ref{fig:africa}.
The empirical covariance function for this variable displays a regular decrease
 and the exponential covariance $C(h) = \alpha_0^{-1} \exp(-\alpha_G |h|)$
provides a reasonably good fit  as shown in Figure \ref{fig:cov}.
Since the decrease of the empirical covariance is approximately linear, one may want
to use a function of the form 
$C(h) = \alpha_0^{-1}  (1-\alpha_G|h|)_+$, 
where $(a)_+$ denotes positive part of $a$, that is
$ C(h) = \alpha_0^{-1} \left ( 1 - \alpha_G |h|\right )$ 
whenever $\quad |h| < \frac 1{\alpha_G}$
and $0$ elsewhere. This covariance is known as the triangle model in
the Geostatistics literature. 
This function provides
visually an even better fit (Fig. \ref{fig:cov}).
Unfortunately, this covariance is valid in one dimension but not in
two dimensions \citep{Chiles99}, which has consequences illustrated
below. 
Using the exponential covariance as a covariance model for tree
abundance (which is a valid
model in any dimension)  enables us to perform spatial prediction
(Fig. \ref{fig:kriging} top left panel) and to derive an assessment of the error realized by the prediction
known as kriging variance (Fig. \ref{fig:kriging} top right panel). Both maps are well behaved and seem
to make sense ecologically and statistically.
Using the triangle model to compute spatial prediction and kriging
variance does not bring any difficulty computer-wise.
The fact that the triangle function is not positive-definite
shows up in the  kriging variance: the latter displays spatial variation that does not mirror
the location of the sampling sites, it is negative in several areas
(Fig. \ref{fig:kriging} bottom right panel) and takes a minimum of $\sigma^2_K=-5720$.
For short, using the triangle covariance in 2 dimensions leads to non-sensical results.

\clearpage
\section{Validity of the stable and  the BRC models}

In addition to $\alpha_2$, the  models we consider involve three or
four parameters.
Positive-definiteness is, however, not influenced by
$\alpha_0,\alpha_G$ and $\alpha_E$ as long as they are
positive.
Therefore, without loss of generality on the mathematical side,
we assume from now on that $\alpha_0=\alpha_G=\alpha_E=1$.

\subsection{Euclidean distance}
If $h$ is $\|x-x'\| = \sqrt{(x_1-x_1')^2+\ldots +  (x_d-x_d')^2}$
(the Euclidean distances in $\rd$)  the stable covariance
$K(h) =  \exp\left[ -   h^{\alpha_2}\right]$
is a valid covariance model in  $\rd$   if and only if $\alpha_2 \in
[0,2]$. Arguments proving these results are given by
\citet{Schoenberg38}.

\noindent
For $u$ defined as $|e-e'|$, the BRC model defined by $C(h,u) =
 \exp\left[ - \left(  h +u \right)^{\alpha_2}\right]$ is a valid covariance
model on $\rd \times \real$ if and only if $\alpha_2 \in
[0,1]$. We give a proof of this original result in the Appendix.

\subsection{Geodesic distance}
We denote by $\sphere^{d-1}$
the unit sphere in $\rd$
and define now $h$ as $\arccos\Big(\sum_{i=1}^d x_ix_i'\Big)$
(geodesic or great circle distance on the
sphere)  while keeping  $u = |e-e'|$.
The stable model is a valid covariance model in  $\sphere^{d-1}$
if and only if $\alpha_2 \in [0,1]$.
Arguments proving this result are given by \citet{Gneiting13}.

\noindent
For the general BRC model on $\sphere^{d-1} \times \real $, we found counter-examples  showing that
for $\alpha_2 =1.001$, the  model is not valid.
Using a continuity argument, this means that no
model with $\alpha_2\geq 1.001$ will be valid.
An instance is
as follows: we consider three points on the sphere with (Lon,Lat) coordinates
$x_1=(-60.0,60)$, $x_2=(-60.1,60)$, $x_3=(-60.2,60)$
and  values $e_1=0.1$, $e_2=0.2$, $e_3=0.3$ of an
environmental variable.
We also set $\alpha_0=1$, $\alpha_G=\alpha_e=1/300$, and
$\alpha_2=1.01$.
Under the BRC model the covariance matrix associated to this
configuration is a $9\times 9$
matrix whose minimum eigenvalue is approximately $-1.84 \times
10^{-5}$,
which shows that the matrix is not positive-definite.
A general theoretical result similar to the case of Euclidean distances is still lacking, but we conjecture that the BRC
model on $\sphere^{d-1} \times \real $ is valid if and only if $\alpha_2 \in [0,1]$.

\subsection{Other distances in the plan or the sphere}
It is a common practice in ecology to measure distances in terms of
cumulative cost for an individual to move from a geographical location
to another. This is referred to as cost or resistance distance. There is
considerable flexibility in the way such a distance can be obtained and
the validity of the BRC model should be checked on a case by case
basis.
From the previous paragraphs, it is clear that the choice
of the distance is not innocuous and that a distance that makes sense
ecologically may not lead to a model that is well behaved
mathematically.
We note also that if the cost distance is obtained via numerical
values (without a mathematical expression), there is little
hope for proving the validity of a covariance model as this would
involve checking all possible sums of the form given in
Equation~\eqref{eq:var_cl}.

\clearpage
\section{Alternate covariance models for applications in
  evolutionary biology}

\subsection{Valid gluing of  the Euclidean geographical distance and the environmental distances}
If  the distance on $\rd \times
\real$ is defined as
\begin{equation}
d[(x,e),(x',e')] = \sqrt{\sum_{i=1}^d  (x_i-x_i')^2 +  (e-e')^2}
\end{equation}
then any  valid covariance model  on $\rd \times
\real$  can be used.
In
particular, $\exp(-d^{\; \alpha_2})$ is a valid model for $\alpha_2 \in
(0,2]$.
See classical textbooks by \citet{Chiles99} and
\citet{Diggle07} for alternative choices.
With a valid model in hands, quantifying the relative effect of distance
and environment variables as suggested by \citet{Bradburd13}
can be done by re-scaling the distance as $\sqrt{\sum_{i=1}^d  \alpha_G(x_i-x_i')^2 +  \alpha_E(e-e')^2}$.

For data gathered at large scale, one has to use geographic distances on the
sphere and there seems to be no straightforward way to combine the geodesic distance
with the environmental distance along this line to obtain a valid model.

\subsubsection{Sums and products of valid models}
If $C_G(h)$ is a valid model on $\rd$ or $\sphere^{d-1}$ and
$C_E(u)$ is a valid model on $\real$, then
\begin{equation}
 C_1(h,u)  = C_G(h) + C_E(u)
\end{equation} and
\begin{equation}
 C_2(h,u)  = C_G(h) \times C_E(u)
\end{equation}
are  valid models for which we give examples in Table~\ref{tab:examples}.

\subsection{Space-time covariance models}
Covariance models developed to handle spatio-temporal
data can be used readily for the analysis of data of the form
considered by \citet{Bradburd13}.
The list of such models on $\rd \times
\real$ or $\sphere^{d-1} \times
\real$ is still limited but  it comes with clear guidelines about
the valid range of parameters.
We refer interested  readers to  recent spatial  statistics books 
\citet{Gelfand10} and \citet{Porcu10}.

\begin{table}
\begin{tabular}{lll}

\hline
\hline

Model name & Covariance function & Parameter range \\

\hline
& & \\
Stable & $C(h)= \exp \left ( -h^{\alpha} \right )$ & $\alpha \in
(0,2]$ on $\real^d$\\
& & $\alpha \in(0,1]$ on $\sphere^{d-1}$\\
& & \\
\hline
& & \\
BRC & $C(h,u)= \exp \left ( -(h + u)^{\alpha} \right )$ & $\alpha \in
(0,1]$ on $\real^d\times \real$\\
& & Unknown for $\sphere^{d-1}\times \real$\\
& & \\
\hline
& & \\
Modified BRC & $C((x,e),(x',e'))= $&\\
&$\exp \left ( -\sqrt{\sum_{i=1}^d (x_i-x_i')^2 + (e-e')^2}^{\;\alpha} \right )$ & $\alpha \in
(0,2]$ on $\real^d\times \real$\\
& & \\
\hline
& & \\
Sum of stable   & $C(h,u)= \exp \left ( -h^{\alpha})+ \exp(-
u^{\beta}\right )$ & $(\alpha,\beta) \in (0,2] \times (0,2]$ on
    $\real^d\times \real$ \\
models& &  $(\alpha,\beta) \in (0,1] \times (0,2]$ on
    $\sphere^{d-1}\times \real$\\
& & \\
\hline
& & \\
Product of stable   & $C(h,u)= \exp \left ( -h^{\alpha}) \times \exp(-
u^{\beta}\right )$ & $(\alpha,\beta) \in (0,2] \times (0,2]$ on
    $\real^d\times \real$ \\
models& &  $(\alpha,\beta) \in (0,1] \times (0,2]$ on $\sphere^{d-1}\times \real$\\
& & \\
\hline
\end{tabular}
\caption{Summary of covariance models with range of validity.
In the table,  $u$ is the environmental distance $|e-e'|$
while $h$ refers to the Euclidean distance $\|x-x'\| = \sqrt{
    \sum_{i=1}^d (x_i - x_i')^2} $ on $\real^d$,
and to the geodesic distance $\arccos\Big(\sum_{i=1}^d x_ix_i'\Big)$ on the unit sphere $\sphere^{d-1}$ of $\rd$.}\label{tab:examples}
\end{table}

\section{Conclusion}

There are limitations on the parameter range for the stable and
the BRC models and they depend on the way distances are measured.
We provide clear guidelines for the case of Euclidean
distances while the case of geodesic distances still requires more
work. For cost distances, a general theoretical statement is not possible
and checking the validity for numerically-derived distances seems out
of reach. We recommend users to be cautious when using cost distances
in this context.
These limitations have remained un-noticed so far and some of the
earlier works making use of these models have been based on invalid
parameter ranges. However, in agreement with our findings, none of
these  earlier studies reported empirically estimated values outside
the valid ranges we establish.
Our work provides some guidelines to update corresponding programs and 
we are happy to note that they are currently used 
to  update the BEDASSLE computer program  (G. Bradbrud, personal communication).

\clearpage
\section{Funding}
E.P. is funded by Proyecto Fondecyt Regular n. 1130647, M.B. by
Proyecto Fondecyt Iniciaci\'on n. 11121408, 
G.G by Agence Nationale de la Recherche project ANR-09-BLAN-0145-01 
and the Danish e-Infrastructure Cooperation.



\appendix
\clearpage
\section{Appendix: valid parameter range for the BRC model}

We determine here for which values of $\alpha_2$ the function from Equation~\eqref{eq:BRC} is a covariance function. A map $\gamma$ from $\real^d \times \real$ into $\real$ is called a \emph{variogram} if it represents the variance of the increments of an intrinsically stationary random field, i.e.
$$
    \gamma \left (x_j-x_i, e_j-e_i\right ) = \Var \left ( Z(x_j,e_j) - Z(x_i,e_i) \right ).
$$
Variograms are real-valued \emph{negative definite functions}, i.e.\ for any finite family of
points $\{(x_i,e_i)\}_{i=1}^N$ and constants $\{a_i\}_{i=1}^N$ with $\sum_{i=1}^N a_i=0$, we have
$$
    \sum_{i=1}^N \sum_{j=1}^N \gamma \left ( x_j-x_i, e_j-e_i \right ) a_i a_j \le 0.
$$
The connection between variograms and covariance functions is due to \citet{Schoenberg38}:
$C:\rd\times\real\to\real$ is a covariance function if and only if
$C(x,e) = \exp(-r\gamma(x,e))$ where $\gamma(x,e)$ is a variogram.

Thus, we can re-cast the question about the valid range of parameter in the following way:
\begin{equation}\label{leo_messi}
    \text{for which $\alpha_2>0$ is the function $(h,u)\mapsto\left ( h + u \right )^{\alpha_2}$ a variogram?}
\end{equation}
As before, $h = \|x\| = \sqrt{x_1^2 + \ldots + x_d^2}$ is the Euclidean distance (taken from the origin) in $\rd$ and $u=|e|$ is the ecological distance (in $\real$, also relative to the origin). In order to simplify the notation, we write $\alpha$ instead of $\alpha_2$.

It is known that every \emph{continuous} variogram on $\rn$ is given by a \emph{L\'evy--Khintchine formula}:
\begin{equation}\label{lkf}
    \gamma(\eta)
    = \frac 12 \eta\cdot Q\eta + \int_{y\neq 0} \Big(1-\cos \Big({\textstyle\sum\limits_{i=1}^n \eta_i y_i}\Big)\Big)\,\nu(dy),\quad\eta\in\rn,
\end{equation}
where $Q$ is a symmetric positive semi-definite $n\times n$ matrix, and $\nu$ is a measure on $\rn\setminus\{0\}$ such that $\int_{y\neq 0}\|y\|^2/(1+\|y\|^2)\,\nu(dy)<\infty$; $\gamma$ is uniquely determined by $(Q,\nu)$ and vice versa. Typical examples of continuous variograms on $\rn$ are
$$
    \|\eta\|^2,\quad
    \eta\cdot Q\eta,\quad
    1-\cos y\cdot\eta,\quad
    \log(1+\|\eta\|^2),\quad
    \|\eta\|^\alpha\; (0<\alpha<2).
$$
A good source for variograms (which are also known as negative
definite real functions) are the monographs by \citet{Berg75} and \citet{Schilling12}. We only need the following properties.
\begin{description}
\item[(A)] Subadditivity:
    If $\gamma(\eta)$ is a continuous variogram, then $\sqrt{\gamma(\eta+\eta)} \leq \sqrt{\gamma(\eta)} + \sqrt{\gamma(\eta)}$. In particular, $\gamma(\eta)$ grows at most like $\|\eta\|^2$ as $\|\eta\|\to\infty$.
\item[(B)] Closure under pointwise limits:
    If $\gamma_j(\eta), j=1,2,\ldots$ are continuous variograms such that the limit $\gamma(\eta):=\lim_{j\to\infty}\gamma_j(\eta)$ exists and is continuous, then $\gamma(\eta)$ is a continuous variogram.
\item[(C)]
    Let $\eta\mapsto \gamma(\eta)$ be a continuous variogram on $\real^d$ and write $\eta = (\eta',\eta'')$ where $\eta'\in\rn$, $\eta''\in\real^{d-n}$. Then $\eta'\mapsto \gamma(\eta',0)$ is a continuous variogram on $\rn$.
\item[(D)]
    Let $\gamma(\eta')$, $\psi(\eta'')$ be continuous variograms on $\rn$ and $\real^m$, respectively. Then $(\eta',\eta'')\mapsto \gamma(\eta')+\psi(\eta'')$ is a continuous variogram on $\rd = \real^{n+m}$.
\end{description}
The variogram property is also preserved under a technique called \emph{Bochner's subordination}, cf.\ \cite{Schilling12}. At the level of the random variables this corresponds to a mixture of the processes with a further infinitely divisible random variable, at the level of variograms this is just a composition with the class of so-called \emph{Bernstein functions}. These are also given by a L\'evy--Khintchine formula
$$
    f(\lambda) = b\lambda + \int_{0+}^{\infty} (1-e^{-s\lambda})\,\mu(ds), \quad \lambda\geq 0,
$$
where $b\geq 0$ and $\mu$ is a measure on $(0,\infty)$ such that $\int_0^\infty s(1+s)^{-1}\,\mu(ds)<\infty$. Typical examples of Bernstein functions are
$$
    \lambda,\quad
    \lambda^\alpha\; (0<\alpha< 1),\quad
    \log(1+\lambda).
$$
\begin{theorem}
    If $\gamma(\eta)$ is a continuous variogram and $f$ is a Bernstein function, then $f(\gamma(\eta))$ is again a continuous variogram.
\end{theorem}

We now have all ingredients for the
\begin{proof}[Proof of the valid parameter range]
Note that $\psi(\eta) = \|\eta\| = \sqrt{\eta_1^2 + \ldots +\eta_d^2}$ and $\phi(\tau) = |\tau|$ are continuous variograms in $\real^d$ and $\real$, respectively. Moreover, take the Bernstein function $f(\lambda)=\lambda^\alpha$, $\lambda>0$; the corresponding mixing random variables are one-sided $\alpha$-stable random variables (if $0<\alpha < 1$) or a deterministic drift (if $\alpha=1$). By property \textbf{(D)} and subordination,
\begin{equation} \label{donald_duck}
    (\eta,\tau) \mapsto \gamma_{\alpha} (\eta,\tau) := (\|\eta\|+|\tau|)^\alpha,\quad 0<\alpha\leq 1,
\end{equation}
is a continuous variogram.

On the other hand, by the quadratic growth property, see \textbf{(A)}, it is clear that $\gamma_\alpha(\eta,\tau)$ is not a variogram if $\alpha>2$.

Let us now consider the case where $\alpha\in (1,2]$. Assume first that $\alpha=2$. Then
$$
     (\|\eta\|+|\tau|)^2 = \|\eta\|^2 + 2\,\|\eta\|\cdot|\tau| + \tau^2.
$$
Since $\|\eta\|^2+\tau^2$ would appear in the L\'evy--Khintchine formula \eqref{lkf} as part of the expression involving the matrix $Q$, it is enough to prove or disprove that the mixed term $c(\eta,\tau) := \|\eta\|\cdot|\tau|$ is a continuous variogram. But
$$
    \sqrt{\|\eta\|\cdot|\tau|} = \sqrt{c(\eta,\tau)} \geq \sqrt{c(\eta,0)} + \sqrt{c(0,\tau)} = 0,
$$
which means that $\sqrt{c(\eta,\tau)}$ is \emph{not} sub-additive, violating the subadditivity property \textbf{(A)}, i.e.\
$$
    (\eta,\tau)\mapsto (\|\eta\|+|\tau|)^2
    \quad\text{is not a variogram}.
$$
Now we use the property \textbf{(B)}:
Clearly, $\lim_{j\to\infty} (\|\eta\|+|\tau|)^{2-1/j} = (\|\eta\|+|\tau|)^2$. Since variograms are preserved under pointwise limits, we conclude from this, and the subordination argument, that there is some $1\leq b < 2$ such that
$$
    (\eta,\tau)\mapsto (\|\eta\|+|\tau|)^\alpha\quad
    \text{is\ \ } \begin{cases}
    \text{\ a continuous variogram if} &0<\alpha\leq b\\
    \text{\ not a continuous variogram if} &\alpha > b.
    \end{cases}
$$
We conclude the proof by showing that necessarily $b=1$. Use Property \textbf{(C)} above, and suppose that the function in Equation~\eqref{donald_duck} is a variogram on $\real^d$. Then the function
$$
    \tilde{\gamma}(\eta_1,\tau)
    := \gamma_{\alpha} \left ( (\eta_1,0,\ldots,0),\tau \right )
$$
is a variogram on $\real \times \real$. Arguments by \citet{Zastavnyi00} show that this is true if and only if $\alpha \le 1$, which completes the proof.
\end{proof}

\end{document}